\documentclass[acmsmall,nonacm]{acmart}

\usepackage{amsthm}
\usepackage[binary-units=true]{siunitx}
\usepackage{amsmath}

\newcommand*{\symDefine}[2]{\newcommand{#1}{#2}}

\symDefine{\symActiveIdx}{A}
\symDefine{\symBucketIdx}{b}
\symDefine{\symRandomBit}{X}
\symDefine{\symRandomBitAll}{U}
\symDefine{\symSubBits}{u}
\symDefine{\symSuccessProbability}{p}
\symDefine{\symSelectIdx}{s}
\symDefine{\symGammaVar}{G}
\symDefine{\symSomeIdx}{i}
\symDefine{\symNonZeroIndex}{\symSomeIdx}
\symDefine{\symKey}{k}
\symDefine{\symIntervalIdx}{m}
\symDefine{\symIntervalIdxMax}{M}
\symDefine{\symNumBuckets}{n}
\symDefine{\symNumBucketsMax}{N}
\symDefine{\symRndVal}{R}
\symDefine{\symActiveIdxIdx}{l}
\symDefine{\symNonZero}{\symIntervalIdx}
\symDefine{\symIntervalIdxLow}{{\symNonZero_0}}
\symDefine{\symIntervalIdxOne}{{\symNonZero_1}}
\symDefine{\symIntervalIdxTwo}{{\symNonZero_2}}
\symDefine{\symUniformVar}{U}
\symDefine{\symIntervalSubIdx}{v}
\symDefine{\symRandomSubIntAll}{W}
\symDefine{\symRandomIntAll}{V}
\symDefine{\symRandomSubInt}{Z}
\symDefine{\symRandomInt}{Y}

\symDefine{\symSomeRandomVariateOne}{K}
\symDefine{\symSomeRandomVariateTwo}{L}

\symDefine{\symNumberRandomValues}{T}
\symDefine{\symNumberRandomValuesJH}{\symNumberRandomValues_\text{JH}}
\symDefine{\symNumberRandomValuesJBH}{\symNumberRandomValues_\text{JBH}}
\symDefine{\symNumberRandomValuesJBHStar}{\symNumberRandomValues_{\text{JBH*}}}

\symDefine{\symProbability}{\Pr}
\symDefine{\symSomeDistribution}{D}

\symDefine{\symSetActiveIndices}{\mathcal{A}}
\symDefine{\symInterval}{\mathcal{I}}
\symDefine{\symComplexity}{\mathcal{O}}
\symDefine{\symPRG}{\mathcal{R}}
\symDefine{\symSomeSet}{\mathcal{S}}

\symDefine{\symDelta}{\Delta}
\symDefine{\symRate}{\lambda}
\symDefine{\symRatio}{\alpha_{\symNumBuckets}}

\DeclareMathOperator*{\symUniform}{Uniform}
\DeclareMathOperator*{\symBernoulli}{Bernoulli}
\DeclareMathOperator*{\symGeometric}{Geom}
\DeclareMathOperator*{\symExpDist}{Exp}
\DeclareMathOperator*{\symGammaDist}{Gamma}
\DeclareMathOperator*{\symConsistentHash}{ch}
\DeclareMathOperator*{\symVariance}{Var}
\DeclareMathOperator*{\symExpectation}{\mathbb{E}}
\DeclareMathOperator*{\argmin}{arg\,min}
\DeclareMathOperator*{\symBitCount}{bitcount}

\newcommand*\bitwiseXor{\oplus}

\usepackage[commandnameprefix=always,final]{changes}
\usepackage{xifthen}
\usepackage[nolist,nohyperlinks]{acronym}
\usepackage[ruled,vlined]{algorithm2e}
\usepackage{mathtools}
\usepackage[capitalise,noabbrev]{cleveref}
\usepackage{enumitem}
\usepackage{courier} 
\usepackage{listings, xcolor}

\definecolor{codegreen}{rgb}{0,0.6,0}

\SetKwProg{Fn}{function}{}{}
\SetKwComment{Comment}{$\triangleright$\ }{}

\SetCommentSty{mycommfont}
\SetKwBlock{Loop}{loop}{end loop}
\SetKw{Break}{break}
\SetKw{Continue}{continue}

\newcommand{\myAlg}[2][]{
 \ifthenelse{\isempty{#1}}%
 {\begin{figure}}
 {\begin{figure}[#1]}
 \begingroup 
 \csname @twocolumnfalse\endcsname
 \noindent
 \resizebox{\columnwidth}{!}{%
 \begin{minipage}{1.2\columnwidth}
 \begin{algorithm}[H]
 \DontPrintSemicolon
 {#2}
 \end{algorithm}
 \end{minipage}%
 }
 \endgroup
 \end{figure}
}

\newcommand{\myListing}[2][]{
 \ifthenelse{\isempty{#1}}%
 {\begin{figure}}
 {\begin{figure}[#1]}
 \begingroup 
 \csname @twocolumnfalse\endcsname
 \noindent
 \resizebox{\columnwidth}{!}{%
 \begin{minipage}{1.2\columnwidth}
 \begin{lstlisting}[language=Java]
 {#2}
 \end{lstlisting}
 \end{minipage}%
 }
 \endgroup
 \end{figure}
}

\begin{acronym}
 \acro{ICWS}{improved consistent weighted sampling}
 \acro{PRG}{pseudorandom generator}
\acro{JMH}{Java Microbenchmark Harness}
\acro{LCG}{linear congruential generator}
\end{acronym}

\allowdisplaybreaks
\begin{document}
\title{JumpBackHash: Say Goodbye to the Modulo Operation to Distribute Keys Uniformly to Buckets}
\author{Otmar Ertl}
\affiliation{
  \institution{Dynatrace Research}
  \city{Linz}
  \country{Austria}
}
\email{otmar.ertl@dynatrace.com}

\begin{abstract}

  \paragraph{Introduction}
  Distributed data processing and storage systems require efficient methods to distribute keys across buckets. While simple and fast, the traditional modulo-based mapping is unstable when the number of buckets changes, leading to spikes in system resource utilization, such as network or database requests. Consistent hash algorithms minimize remappings but are either significantly slower, require floating-point arithmetic, or are based on a family of hash functions rarely available in standard libraries. This work introduces JumpBackHash, a consistent hash algorithm that overcomes those shortcomings.

  \paragraph{Methodology}
  JumpBackHash applies the concept of active indices borrowed from consistent weighted sampling, which inherently leads to consistency. It generates the active indices in reverse order, which avoids floating-point operations, enables the minimization of consumed random values and the use of a standard pseudorandom generator, and finally leads to a very efficient algorithm.

  \paragraph{Results}
  Theoretical analysis shows that JumpBackHash has an expected constant runtime. The expected value and the variance of the number of consumed random values perfectly agree with the experiments. Empirical tests also confirm the consistency.

  \paragraph{Conclusion}
  JumpBackHash offers a fast and efficient solution for uniformly distributing keys across buckets in distributed systems. Its simplicity, performance, and the availability of a production-ready Java implementation as part of the Hash4j open source library make it a viable replacement for the modulo-based approach for improving assignment and system stability.

\end{abstract}


\maketitle
\pagestyle{plain}

\section{Introduction}
Distributed data processing and storage requires a strategy how tasks or data are mapped across available resources. In more abstract terms, the strategy must be able to distribute keys to a given number of buckets. Ideally, the keys are uniformly assigned to the buckets to achieve balanced utilization of resources.
If there are $\symNumBuckets$ buckets labeled $0,1,\ldots,\symNumBuckets-1$, a very popular pattern to map a given integer key $\symKey$ to a bucket index $\symBucketIdx$ uses the modulo operation according to
\begin{equation}
  \label{equ:modulo_pattern}
  \symKey \bmod \symNumBuckets \rightarrow \symBucketIdx.
\end{equation}
If the key $\symKey$ is the result of some high-quality hash function \cite{Yi,Vaneev, Peters} and can be considered like a true random value, and the number of buckets is much smaller than the maximum possible hash value, which is usually the case for 64-bit hash values, \chadded{then} the keys \chadded{can be assumed to} be uniformly distributed over the buckets \chadded{in practice}, \chreplaced{thereby satisfying}{such that}
\begin{equation}
  \label{equ:modulo_uniformity}
  \symProbability((\symKey \bmod \symNumBuckets) = \symBucketIdx) = \frac{1}{\symNumBuckets}.
\end{equation}

Unfortunately, a change in the number of buckets will also change the mapping of nearly all keys. If $\symNumBuckets$ is increased to $\symNumBuckets+1$, every key will be mapped to a different bucket with a probability of $\frac{\symNumBuckets}{\symNumBuckets+1}$ which approaches 1 for large $\symNumBuckets$. The same is true if the number of buckets is reduced. All those reassignments would happen at the same time and cause pressure on a distributed system. For example, this might lead to timely concentrated higher network load as data needs to be shifted to other physical instances. Or, this could also result in a short-term increase of database requests to load key-specific settings or missing cache entries, which in turn may cause back pressure on data streams.

Therefore, it is critical to minimize the number of reassignments while preserving a uniform distribution.
To maintain uniformity after increasing the number of buckets from $\symNumBuckets$ to $\symNumBuckets+1$, some keys must be shifted from the already existing $\symNumBuckets$ buckets to the newly added $(\symNumBuckets+1)$-th bucket with index $\symNumBuckets$.
Since the expected proportion of keys in the new bucket should be $\frac{1}{\symNumBuckets+1}$ at the end, it is obvious that on average at least $\frac{1}{\symNumBuckets+1}$ of the keys must switch buckets.
Consistent hash algorithms achieve this lower bound and thus minimize the number of reassignments while mapping keys uniformly to a given number of buckets. For them, the probability that a key gets reassigned would be close to 0 for large $\symNumBuckets$, because $\frac{1}{\symNumBuckets+1}\rightarrow 0$ as $\symNumBuckets\rightarrow \infty$.

\chreplaced
{
  If a consistent hash function also had a comparable evaluation time, it would be generally preferable to the modulo approach. Pattern \eqref{equ:modulo_pattern} could then be replaced by
}
{
  If there was no performance difference, it would be generally wise to replace pattern \eqref{equ:modulo_pattern} by
}
\begin{equation*}
  \symConsistentHash(\symKey, \symNumBuckets) \rightarrow \symBucketIdx
\end{equation*}
where $\symConsistentHash(\symKey, \symNumBuckets)$ denotes \chreplaced{the}{a} consistent hash function.

\begin{figure}[t]
  \centering
  \includegraphics[width=0.8\linewidth]{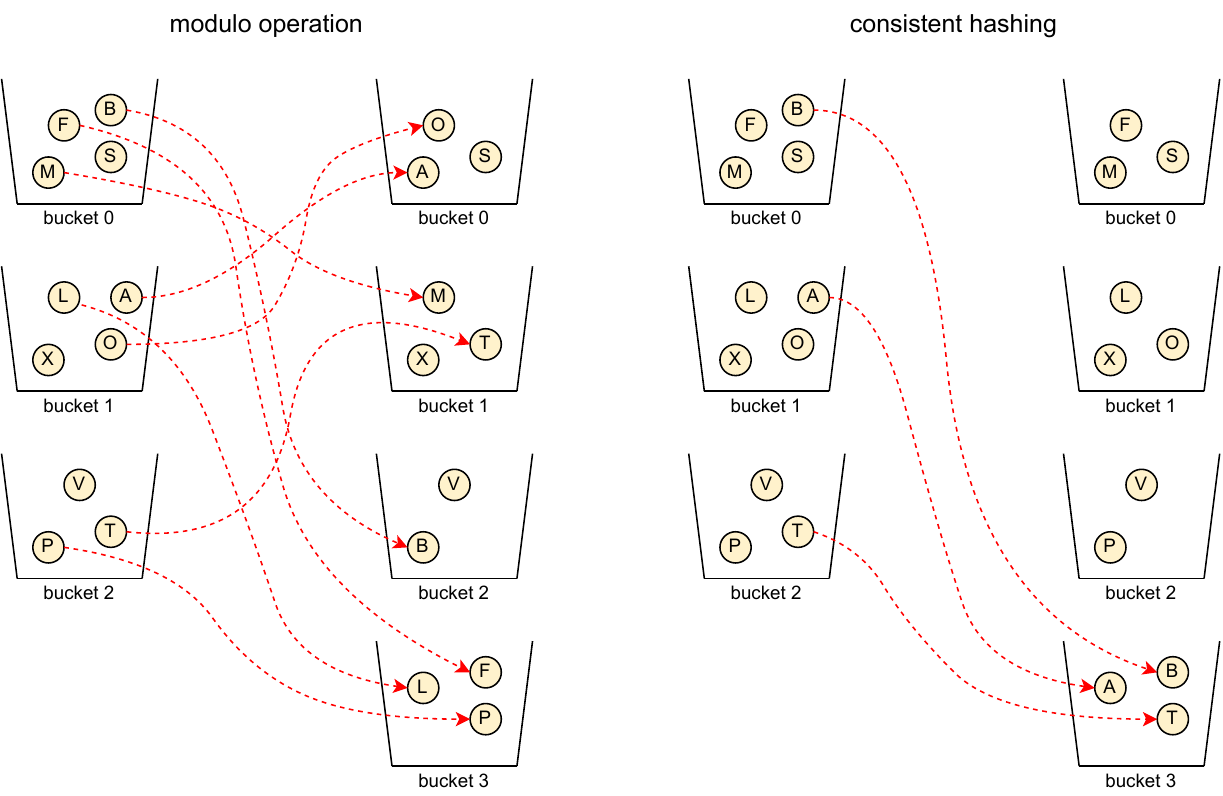}
  \caption{When changing the number of buckets, some keys (letters) must be mapped to different buckets to restore balance. In contrast to the modulo operation, which puts almost all keys into different buckets, consistent hashing minimizes the expected number of reassignments (arrows).}
  \label{fig:assignment}
\end{figure}

\cref{fig:assignment} illustrates the difference between the modulo-based approach and consistent hashing in terms of the number of reassignments when the number of buckets changes from 3 to 4. This work presents JumpBackHash, which is a simple consistent hash function with a performance comparable to the modulo-based approach \eqref{equ:modulo_pattern}.
\cref{tab:notation} provides an overview of the used notations.

\begin{table}[t]
  \caption{Notations.}
  \label{tab:notation}
  \footnotesize
  \begin{tabular*}{\linewidth}{@{\extracolsep{\fill}} ll }
    \toprule
    Symbol
    &
    Description
    \\
    \midrule
    $\lfloor\ldots\rfloor$
    &
    floor function, e.g. $\lfloor 3.7\rfloor = 3$
    \\
    $\langle\ldots\rangle_2$
    &
    binary representation, e.g. $\langle 110\rangle_2 = 6$
    \\
    $\bitwiseXor$
    &
    bitwise XOR, e.g. $\langle 1010\rangle_2 \bitwiseXor \langle 1100\rangle_2 = \langle 0110\rangle_2$
    \\
    $\symBitCount$
    &
    number of set bits, e.g. $\symBitCount(\langle 10110\rangle_2)=3$
    \\
    $\symKey$
    &
    key, typically a 64-bit hash value
    \\
    $\symNumBuckets$
    &
    number of buckets, $\symNumBuckets>0$
    \\
    $\symNumBucketsMax$
    &
    some constant upper bound such that $\symNumBuckets\leq \symNumBucketsMax$ holds, see \cref{sec:jumping_backwards}
    \\
    $\symIntervalIdxMax$
    &
    some constant such that $\symNumBuckets\leq 2^\symIntervalIdxMax$ holds, see \cref{sec:improved_active_index_generation}
    \\
    $\symConsistentHash(\symKey,\symNumBuckets)$
    &
    consistent hash function, see \cref{sec:consistent_hash_function}
    \\
    $\symSetActiveIndices$
    &
    set of active indices, compare \cref{sec:active_indices}
    \\
    $\symUniform(\symSomeSet)$
    &
    uniform distribution over the set $\symSomeSet$
    \\
    $\symExpDist(\symRate)$
    &
    exponential distribution with rate parameter $\symRate$
    \\
    $\symGeometric(\symSuccessProbability)$
    &
    geometric distribution with success probability $\symSuccessProbability$ and positive support
    \\
    $\symBernoulli(\symSuccessProbability)$
    &
    Bernoulli distribution with probability $\symSuccessProbability$
    \\
    $\sim$
    &
    is distributed as, e.g. $\symBernoulli(0.5)\sim\symUniform(\lbrace 0, 1\rbrace)$
    \\
    $\symProbability$
    &
    probability, e.g. $\symProbability(\symBernoulli(0.3) = 1) = 0.3$
    \\
    $\symExpectation$
    &
    \chreplaced{expected value (mean)}{mean}, e.g. $\symExpectation(\symUniform(\lbrace 0, 1\rbrace)) = 0.5$
    \\
    $\symVariance$
    &
    variance
    \\
    $\symPRG$
    &
    \acf*{PRG}
    \\
    $\symPRG[\symSomeDistribution]$
    &
    sample from distribution $\symSomeDistribution$ using $\symPRG$
    \\
    $\llbracket\symNumBuckets\rrbracket$
    &
    $\llbracket\symNumBuckets\rrbracket:= \lbrace 0,1,2,\ldots,\symNumBuckets-1\rbrace$
    \\
    $\symInterval_\symIntervalIdx$
    &
    $\symInterval_\symIntervalIdx:= \lbrace 2^{\symIntervalIdx},2^{\symIntervalIdx}+1,\ldots, 2^{\symIntervalIdx+1}- 1\rbrace = \llbracket2^{\symIntervalIdx+1}\rrbracket \setminus \llbracket2^{\symIntervalIdx}\rrbracket$, $\symIntervalIdx \geq 0$
    \\
    $\symComplexity$
    &
    big O notation, e.g. $\sum_{\symSomeIdx=1}^{\symNumBuckets}\frac{1}{\symSomeIdx}=\symComplexity(\ln \symNumBuckets)$
    \\
    $\symRatio$
    &
    $\symRatio := \frac{1}{\symNumBuckets} 2^{\lfloor \log_2(\symNumBuckets-1)\rfloor+1}\in[1,2)$, $\symNumBuckets\geq 2$
    \\
    \bottomrule
  \end{tabular*}
\end{table}

\subsection{Consistent Hash Functions}
\label{sec:consistent_hash_function}
A \chreplaced{\textit{consistent hash function}}{consistent hash function,} is a deterministic function that maps a key $\symKey$ to a bucket index from $\llbracket\symNumBuckets\rrbracket := \lbrace 0,1,2,\ldots,\symNumBuckets-1\rbrace$, where $\symNumBuckets>0$ is the number of buckets, and that has the following two properties:

\begin{description}
  \item[Uniformity:] The probability that a random key is assigned to a bucket is the same for all $\symNumBuckets$ buckets which can be written as \cite{Ioffe2010,Leu2023,Manasse2010}
    \begin{equation*}
      \symProbability(\symConsistentHash(\symKey, \symNumBuckets) = \symBucketIdx) = \frac{1}{\symNumBuckets}.
    \end{equation*}
    \chadded{As before in \eqref{equ:modulo_uniformity}, this condition requires the universe of keys to be infinite or at least much larger than $\symNumBuckets$, which is fulfilled in practice, for example, when $\symKey$ is a 64-bit hash value and $\symNumBuckets$ a 32-bit integer.}
    Uniform hash functions are sometimes also called balanced \cite{Dong2023,Mendelson2021,Coluzzi2023a,Coluzzi2023} or regular \cite{Masson2024}.
  \item [Monotonicity:] If the number of buckets is increased \chadded{by 1} and a key is mapped to a different bucket, the bucket is always the last newly added one \cite{Coluzzi2023, Coluzzi2023a,Dong2023}, which can be expressed as
        \begin{equation}
          \label{equ:def_mononicity}
          \symConsistentHash(\symKey, \symNumBuckets) \neq \symConsistentHash(\symKey, \symNumBuckets+1) \Rightarrow \symConsistentHash(\symKey, \symNumBuckets+1) = \symNumBuckets.
        \end{equation}
        This property is sometimes also referred to as minimal disruption \cite{Mendelson2021} or consistency \cite{Ioffe2010}. In contrast, in this paper, we use the term consistency to describe the combination of monotonicity and uniformity.
\end{description}

While the modulo-based approach is uniform according to \eqref{equ:modulo_uniformity}, it is not monotonic. A minimal example violating \eqref{equ:def_mononicity} is given by $\symKey = 4$ and $\symNumBuckets=2$ when $\symConsistentHash(\symKey, \symNumBuckets) := \symKey\bmod \symNumBuckets$.
The implication in \eqref{equ:def_mononicity} is actually an equivalence, because $\symConsistentHash(\symKey, \symNumBuckets+1) = \symNumBuckets > \symConsistentHash(\symKey, \symNumBuckets)\in\llbracket \symNumBuckets \rrbracket$. Therefore, the probability of reassignments for consistent hash functions is given by
$\symProbability(\symConsistentHash(\symKey, \symNumBuckets) \neq \symConsistentHash(\symKey, \symNumBuckets+1)) = \symProbability(\symConsistentHash(\symKey, \symNumBuckets+1) = \symNumBuckets) = \frac{1}{\symNumBuckets+1}$. Hence, uniformity and monotonicity are sufficient to minimize the expected number of reassignments.

\subsection{Related Work}

Interestingly, consistent hashing can be seen as a special case of consistent weighted sampling \cite{Manasse2010}, where the weight vector has a single dimension with an integer weight equal to the number of buckets $\symNumBuckets$. The adaptation of the \ac{ICWS} algorithm \cite{Ioffe2010} to our problem results in \cref{alg:icws} which obviously has a constant time complexity.
Unfortunately, the algorithm involves expensive floating-point operations including an exponentiation and a logarithm evaluation, which makes it rather slow, especially on embedded or low-power devices without floating-point units.
Nevertheless, the algorithm can be easily implemented, as it relies on an ordinary \ac{PRG} that is seeded with the given key $\symKey$. Standard libraries often already come with modern and reliable \ac{PRG} implementations, such as SplitMix \cite{Steele2014} in Java, that pass a series of statistical tests \cite{TestingRNG}. These libraries sometimes even offer functions for the efficient generation of exponentially distributed random values, as required by \ac{ICWS}.
Although \ac{ICWS} has been around for some time, it was only recently discovered as a consistent hash function when it was added to version 0.14.0 of our open source Hash4j library \cite{Hash4j}.

JumpHash \cite{Lamping2014}, on the other hand, became much more popular, although it was invented a few years afterwards.
It is shown in \cref{alg:jumphash} and will be discussed in more detail later. \chreplaced{Like}{As} \ac{ICWS}, it uses a standard \ac{PRG} and floating-point arithmetic, but avoids expensive transcendental functions. Therefore, it is usually faster than \ac{ICWS}, except for large $\symNumBuckets$ due to its logarithmic time complexity.
JumpHash is widely used in the meantime. For example, it is offered as hash function in the Guava Java library \cite{Guava} or in ClickHouse \cite{ClickHouse}. It is also used by Booking.com for their customer review system \cite{Hiltpolt2022}, by Grafana for trace and metric data distribution \cite{Agarwal2020,Boucault2020}, by VMware Greenplum since version 6 to map data values to segments \cite{Greenplum}, by RabbitMQ since version 3.7.8 in its Consistent Hashing Exchange Plugin for message distribution \cite{RabbitMQ}, or for distributed processing of vehicle license plates \cite{Wang2023a}.

Recently, PowerHash was introduced as the first consistent hash algorithm with constant runtime able to beat JumpHash over the whole range of bucket numbers \cite{Leu2023}. Unlike JumpHash and \ac{ICWS}, it relies on a family of hash functions that are less common in standard libraries and often have to be self-implemented.
Since families of hash functions are used less frequently in practice, their statistical properties such as independence and uniformity are usually less tested and therefore less trustworthy than those of \acp{PRG}.
A family of hash functions is also used by the even more recent FlipHash algorithm \cite{Masson2024}, which is similar to PowerHash but avoids floating-point operations.

\myAlg{
  \caption{Specialization of \acf*{ICWS} \cite{Ioffe2010} to map a key $\symKey$ consistently to $\llbracket\symNumBuckets\rrbracket:= \lbrace 0,1,2,\ldots,\symNumBuckets-1\rbrace$.}
  \label{alg:icws}
  \SetKwFunction{FuncICWS}{\acs*{ICWS}}
  \Fn{\FuncICWS{$\symKey$, $\symNumBuckets$}}{
    $\symPRG\gets$ initialize \acs*{PRG} with seed $\symKey$ \;
    $\symUniformVar\gets\symPRG[\symUniform([0,1))]$\;
    $\symGammaVar\gets\symPRG[\symExpDist(1)]+\symPRG[\symExpDist(1)]$\Comment*[r]{$\symGammaVar$ follows a $\symGammaDist(2,1)$ distribution}
    \Return{$\min(\lfloor\exp(\symGammaVar\cdot(\lfloor\ln(\symNumBuckets)/\symGammaVar + \symUniformVar \rfloor - \symUniformVar))\rfloor, \symNumBuckets -1)$}
    \Comment*[r]{truncate at $\symNumBuckets -1$ to protect against numerical errors}
  }
}

\myAlg{
  \caption{The JumpHash algorithm \cite{Lamping2014} consistently maps $\symKey$ to the integer range $\llbracket \symNumBuckets\rrbracket:= \lbrace 0,1,2,\ldots,\symNumBuckets-1\rbrace$. }
  \label{alg:jumphash}
  \SetKwFunction{FuncJumpHash}{JumpHash}
  \Fn{\FuncJumpHash{$\symKey$, $\symNumBuckets$}}{
    $\symPRG\gets$ initialize \acs*{PRG} with seed $\symKey$ \;
    $\symBucketIdx\gets -1$\;
    $\symBucketIdx'\gets 0$\;
    \While{$\symBucketIdx'<\symNumBuckets$}{
      $\symBucketIdx\gets\symBucketIdx'$\;
      $\symUniformVar\gets \symPRG[\symUniform([0,1))]$\;
      $\symBucketIdx'\gets \lfloor (\symBucketIdx + 1) / \symUniformVar \rfloor$\Comment*[r]{determine next larger active index, compare \eqref{equ:jumphash_recursion}}
    }
    \Return{$\symBucketIdx$}\;
  }
}

Consistent hashing as considered in this work, allows removing buckets only in the reverse order as they were added. A lot of research was done on the more general problem, where buckets can be arbitrarily added and removed \cite{Thaler1998, Appleton2015, Karger1997, Dong2023, Eisenbud2016, Nakatani2021, Mendelson2021, Mirrokni2018,Chen2021, Coluzzi2023a,Zeng2022}. However, they are typically significantly slower, require some state, or are not optimal with respect to the number of reassignments. A recently presented algorithm, MementoHash \cite{Coluzzi2023} builds upon JumpHash and therefore demonstrates how any consistent hash algorithm can be extended to support removals in arbitrary order.

\subsection{Our Contributions}
We introduce JumpBackHash, a fast consistent hash algorithm with expected constant runtime, which also works completely without floating-point operations. In contrast to PowerHash \cite{Leu2023} or FlipHash \cite{Masson2024} which require a family of hash functions, JumpBackHash, like JumpHash \cite{Lamping2014}, only uses a standard \acf{PRG}.

To derive JumpBackHash, we use the concept of active indices, which has already been used for consistent weighted sampling \cite{Manasse2010, Ioffe2010}. This new perspective allows a better understanding of JumpHash and leads us after some stepwise optimizations to JumpBackHash. We also perform a theoretical runtime analysis, which is empirically verified together with uniformity and monotonicity.

\section{Methodology}
\label{sec:basic_idea}
A simple way to define a consistent hash function is
\begin{equation}
  \label{equ:basic_consistent}
  \symConsistentHash(\symKey, \symNumBuckets) := \argmin_{\symBucketIdx\in\llbracket\symNumBuckets\rrbracket} \symRndVal_\symBucketIdx,
\end{equation}
where $\symRndVal_0, \symRndVal_1, \symRndVal_2, \ldots$ is a sequence of independent and identically distributed values obtained by a \ac{PRG} with seed $\symKey$ and a support large enough to ignore accidental collisions in practice.
This function is uniform as the random values $\symRndVal_0, \symRndVal_1,\ldots ,\symRndVal_{\symNumBuckets-1}$ have the same chance of being the minimum. It is also monotonic, because if the number of buckets is increased from $\symNumBuckets$ to $\symNumBuckets+1$, the mapping remains stable, except for the case where $\symRndVal_{\symNumBuckets}$ is the new minimum and thus the result changes to $\symNumBuckets$.

Unfortunately, this approach is not very efficient as it takes $\symComplexity(\symNumBuckets)$ time to generate all relevant random values $\symRndVal_\symBucketIdx$ with $\symBucketIdx\in\llbracket\symNumBuckets\rrbracket$. The order in which the random values $\symRndVal_\symBucketIdx$ are drawn using the \ac{PRG} is not important as long as it is well defined and does not depend on $\symNumBuckets$. For example, if we know an upper bound $\symNumBucketsMax$ such that $\symNumBuckets\leq\symNumBucketsMax$ is always satisfied, we could draw the random values in reverse order according to $\symRndVal_{\symNumBucketsMax-1}, \symRndVal_{\symNumBucketsMax-2},\ldots ,\symRndVal_{1},\symRndVal_{0}$. Even though this would lead to a much worse time complexity of $\symComplexity(\symNumBucketsMax)$ if $\symNumBuckets\ll\symNumBucketsMax$, \eqref{equ:basic_consistent} would still be a valid consistent hash function.

\subsection{Active Indices}
\label{sec:active_indices}
To construct a more efficient consistent hash algorithm based on \eqref{equ:basic_consistent}, we use the concept of \emph{active indices} \cite{Manasse2010, Ioffe2010}.
A bucket index $\symBucketIdx$ is an active index if it satisfies
\begin{equation}
  \label{equ:active_index_def}
  \symRndVal_\symBucketIdx = \min_{\symBucketIdx'\in\llbracket\symBucketIdx+1\rrbracket}\symRndVal_{\symBucketIdx'}.
\end{equation}
By definition, bucket index 0 is always an active index. For $\symBucketIdx > 0$ this definition is equivalent to
$\symRndVal_\symBucketIdx < \min_{\symBucketIdx'\in\llbracket\symBucketIdx\rrbracket} \symRndVal_{\symBucketIdx'}$.
According to \eqref{equ:basic_consistent} only active indices are candidates for the mapping. Let $\symSetActiveIndices$ be the set of active indices, then \eqref{equ:basic_consistent} can be simply expressed as
\begin{equation}
  \label{equ:consistent_hash_active_indices}
  \symConsistentHash(\symKey, \symNumBuckets) := \max(\symSetActiveIndices \cap \llbracket \symNumBuckets \rrbracket).
\end{equation}
If we find an equivalent statistical process to generate the active indices in order and directly, without the detour of generating the random values $\symRndVal_{\symBucketIdx}$ for each individual bucket index, we could save a lot of computation time.

\subsection{JumpHash}
\label{sec:jumphash}

Generating active indices in ascending order is exactly what JumpHash does, even though the algorithm was derived and analyzed without the concept of active indices \cite{Lamping2014}.
Given an active index $\symActiveIdx$, the probability that $\symActiveIdx + 1$ is the next larger active index equals the probability that $\symRndVal_{\symActiveIdx + 1}$ is the smallest among the first $\symActiveIdx+2$ random values $\lbrace\symRndVal_\symBucketIdx\rbrace_{\symBucketIdx\in\llbracket\symActiveIdx+2\rrbracket}$ which occurs with a probability of $\frac{1}{\symActiveIdx+2}$. The probability that $\symActiveIdx+2$ is the next active index relative to $\symActiveIdx$ is given by the probability that $\symActiveIdx+1$ is not an active index multiplied by the probability that $\symRndVal_{\symActiveIdx+2}$ is the smallest value, which is therefore
$(1- \frac{1}{\symActiveIdx+2}) \frac{1}{\symActiveIdx+3} = \frac{\symActiveIdx+1}{(\symActiveIdx+2)(\symActiveIdx+3)}$. In general, the probability that
$\symActiveIdx + \symDelta$ is the next active index relative to $\symActiveIdx$ is given by $\chadded{\left(\prod_{\symSomeIdx=2}^{\symDelta}\left(1 - \frac{1}{\symActiveIdx+\symSomeIdx}\right)\right)} \frac{1}{\symActiveIdx+\symDelta+1}=\frac{\symActiveIdx+1}{(\symActiveIdx+\symDelta)(\symActiveIdx+\symDelta+1)}$.

Hence, the active indices can be generated directly and in ascending order using the recursion
\begin{equation}
  \label{equ:jumphash_recursion}
  \symActiveIdx_{\symActiveIdxIdx+1} = \lfloor(\symActiveIdx_{\symActiveIdxIdx} + 1) / \symUniformVar\rfloor \qquad \text{with $\symUniformVar\sim \symUniform([0,1))$ and $\symActiveIdx_0=0$,}
\end{equation}
because $\symActiveIdx_{\symActiveIdxIdx+1} = \symActiveIdx_{\symActiveIdxIdx} + \symDelta$ holds if $\symUniformVar\in (\frac{\symActiveIdx_{\symActiveIdxIdx} + 1}{\symActiveIdx_{\symActiveIdxIdx} + \symDelta+1},\frac{\symActiveIdx_{\symActiveIdxIdx} + 1}{\symActiveIdx_{\symActiveIdxIdx} + \symDelta}]$, which occurs with a probability of $\frac{\symActiveIdx_{\symActiveIdxIdx} + 1}{\symActiveIdx_{\symActiveIdxIdx} + \symDelta} - \frac{\symActiveIdx_{\symActiveIdxIdx} + 1}{\symActiveIdx_{\symActiveIdxIdx} + \symDelta+1} = \frac{\symActiveIdx_{\symActiveIdxIdx} + 1}{(\symActiveIdx_{\symActiveIdxIdx} + \symDelta)(\symActiveIdx_{\symActiveIdxIdx} + \symDelta+1)}$ as required.
  The recursion together with \eqref{equ:consistent_hash_active_indices} leads directly to the JumpHash algorithm as shown in \cref{alg:jumphash}.

  The runtime is given by the number of consumed random values $\symNumberRandomValuesJH$ which also corresponds to the number of active indices that need to be generated until the first active index greater than or equal to $\symNumBuckets$ is found. As the very first active index is always 0 and does not have to be generated, $\symNumberRandomValuesJH$ is also the same as the number of active indices in $\llbracket \symNumBuckets\rrbracket$.
  The probability that bucket index $\symBucketIdx$ is active is $\frac{1}{\symBucketIdx+1}$, as this is the probability that the associated random value $\symRndVal_\symBucketIdx$ is the smallest among the first $\symBucketIdx+1$ random values $\lbrace\symRndVal_{\symBucketIdx'}\rbrace_{\symBucketIdx'\in\llbracket\symBucketIdx+1\rrbracket}$.
  Therefore, $\symNumberRandomValuesJH$ is distributed as
  \begin{equation*}
    \symNumberRandomValuesJH
    \sim
    \sum_{\symBucketIdx=0}^{\symNumBuckets-1}
    \symBernoulli\!\left(\frac{1}{\symBucketIdx+1}\right)
    \sim
    \sum_{\symSomeIdx=1}^{\symNumBuckets}
    \symBernoulli\!\left(\frac{1}{\symSomeIdx}\right),
  \end{equation*}
  where $\symBernoulli(\symSuccessProbability)$ denotes a Bernoulli random variable with probability $\symSuccessProbability$.
  When using
  \begin{align*}
    \symExpectation\!\left(\sum_{\symSomeIdx=1}^{\symNumBuckets}\symSomeRandomVariateOne_\symSomeIdx\right) & = \sum_{\symSomeIdx=1}^{\symNumBuckets} \symExpectation(\symSomeRandomVariateOne_\symSomeIdx),
    \\
    \symVariance\!\left(\sum_{\symSomeIdx=1}^{\symNumBuckets}\symSomeRandomVariateOne_\symSomeIdx\right)    & = \sum_{\symSomeIdx=1}^{\symNumBuckets}\symVariance(\symSomeRandomVariateOne_\symSomeIdx)
  \end{align*}
  for independent $\symSomeRandomVariateOne_\symSomeIdx$
  together with the known expressions for Bernoulli distributed random variables
  \begin{align}
    \label{equ:bernoulli_expectation}
    \symExpectation(\symBernoulli(\symSuccessProbability)) & =\symSuccessProbability,
    \\
    \label{equ:bernoulli_variance}
    \symVariance(\symBernoulli(\symSuccessProbability))    & =\symSuccessProbability(1-\symSuccessProbability),
  \end{align}
  the \chreplaced{expected value}{mean} and the variance of $\symNumberRandomValuesJH$ can be derived as
  \begin{alignat}{3}
    \label{equ:jh_mean}
    \symExpectation
     &
    (\symNumberRandomValuesJH)
     &   &
    =
    \sum_{\symSomeIdx=1}^{\symNumBuckets}\frac{1}{\symSomeIdx}
     &   &
    \leq 1 + \ln\symNumBuckets,
    \\
    \label{equ:jh_var}
    \symVariance
     &
    (\symNumberRandomValuesJH)
     &   &
    =
    \sum_{\symSomeIdx=1}^{\symNumBuckets}\frac{1}{\symSomeIdx}\left(1-\frac{1}{\symSomeIdx}\right)
    =
    \left(\sum_{\symSomeIdx=1}^{\symNumBuckets}\frac{1}{\symSomeIdx}\right)-\left(\sum_{\symSomeIdx=1}^{\symNumBuckets}\frac{1}{\symSomeIdx^2}\right)
     &   &
    \leq
    \ln \symNumBuckets.
  \end{alignat}
  This also shows the logarithmic runtime complexity $\symComplexity(\ln\symNumBuckets)$ of JumpHash.

  \subsection{Jumping Backwards}
  \label{sec:jumping_backwards}
  The way JumpHash generates active indices \eqref{equ:jumphash_recursion} is not unique. If an upper bound $\symNumBucketsMax$ is known for the number of buckets $\symNumBuckets\leq \symNumBucketsMax$, which is always the case in practice as standard integer types are limited, it is also possible to generate active indices in descending order using the recursion
  \begin{equation}
    \label{equ:recursion_backwards}
    \symActiveIdx_{\symActiveIdxIdx+1} \sim \symUniform(\llbracket\symActiveIdx_{\symActiveIdxIdx}\rrbracket) \qquad \text{with $\symActiveIdxIdx\geq 0$ and $\symActiveIdx_0=\symNumBucketsMax$.}
  \end{equation}
  The rationale behind this approach is very simple.
  The next smallest active index is given by the smallest random value from $\lbrace\symRndVal_\symBucketIdx\rbrace_{\symBucketIdx\in\llbracket\symActiveIdx_{\symActiveIdxIdx}\rrbracket}$. As, by definition, the values in this set are all identically distributed, they all have the same chance of being the smallest. As a result, the next smaller active index is uniformly distributed over $\llbracket\symActiveIdx_{\symActiveIdxIdx}\rrbracket$. Based on this recursion \eqref{equ:recursion_backwards} another consistent hash algorithm can be constructed as shown by \cref{alg:jumping_backwards}.

  \myAlg{
    \caption{Consistent hashing by jumping backwards, $\symNumBuckets\leq\symNumBucketsMax$.}
    \label{alg:jumping_backwards}
    \SetKwFunction{FuncJumpingBackwards}{JumpingBackwards}
    \Fn{\FuncJumpingBackwards{$\symKey$, $\symNumBuckets$}}{
      $\symPRG\gets$ initialize random generator with seed $\symKey$\;
      $\symBucketIdx\gets \symNumBucketsMax$\;
      \Repeat{$\symBucketIdx<\symNumBuckets$}{
        $\symBucketIdx\gets \symPRG[\symUniform(\llbracket\symBucketIdx\rrbracket)]$\Comment*[r]{determine next smaller active index, compare \eqref{equ:recursion_backwards}}
      }
      \Return{$\symBucketIdx$}\;
    }
  }

  In contrast to JumpHash, this approach does not involve floating-point operations. However, it requires sampling from an integer range \cite{ONeil2018, Lemire2019}. Another downside is that all active indices in $[\symNumBuckets, \symNumBucketsMax)$ must be generated. If $\symNumBuckets \ll \symNumBucketsMax$, which is usually the case, the runtime complexity would be $\symComplexity(\log \symNumBucketsMax)$, which is worse than the $\symComplexity(\log \symNumBuckets)$ of JumpHash. While we prefer the absence of floating-point operations, we need to make the generation of active indices in descending order much more efficient.

\subsection{Improved Active Index Generation}
\label{sec:improved_active_index_generation}
To construct a more efficient process to generate active indices in descending order we assume without loss of generality that the maximum number of buckets is a power of two $\symNumBucketsMax=2^\symIntervalIdxMax$ and divide the set of bucket indices $\llbracket\symNumBucketsMax\rrbracket$ into disjoint intervals $\symInterval_\symIntervalIdx := \lbrace 2^{\symIntervalIdx},2^{\symIntervalIdx}+1,\ldots, 2^{\symIntervalIdx+1}- 1\rbrace$ according to
\begin{equation*}
  \llbracket\symNumBucketsMax\rrbracket = \lbrace 0 \rbrace \cup \bigcup_{\symIntervalIdx=0}^{\symIntervalIdxMax-1}\symInterval_\symIntervalIdx.
\end{equation*}
A good reason to consider intervals with power-of-two cardinalities is that uniform integer values can be efficiently sampled from $\symInterval_\symIntervalIdx$ using exactly $\symIntervalIdx$ random bits.
The intervals $\symInterval_\symIntervalIdx$ can be ordered as all values of $\symInterval_\symIntervalIdx$ are smaller than that of $\symInterval_{\symIntervalIdx+1}$ and also the cardinality of $\symInterval_\symIntervalIdx$ is smaller than that of $\symInterval_{\symIntervalIdx+1}$. If we compare those intervals in the following, we always refer to this ordering. In particular, this allows us to say that $\symInterval_\symIntervalIdx$ is smaller than $\symInterval_{\symIntervalIdx'}$ if $\symIntervalIdx<\symIntervalIdx'$.

To obtain active indices in descending order, we process the intervals $\symInterval_\symIntervalIdx$ with $\symIntervalIdx\in\llbracket\symIntervalIdxMax\rrbracket$ in descending order and generate all active indices therein also in descending order. The last active index is always 0 by definition \eqref{equ:active_index_def}.
The probability that $\symInterval_{\symIntervalIdx}$ contains an active index is 50\% as the next smaller random value belongs either to $\lbrace \symRndVal_\symBucketIdx\rbrace_{\symBucketIdx\in\llbracket 2^{\symIntervalIdx}\rrbracket}$ or $\lbrace \symRndVal_\symBucketIdx\rbrace_{\symBucketIdx\in\symInterval_{\symIntervalIdx}}$ with the same probability. This decision requires a single random bit $\symRandomBit_\symIntervalIdx\in\lbrace0,1\rbrace$.
If $\symRandomBit_\symIntervalIdx=0$, which occurs in 50\% of all cases, we know that there is no active index in $\symInterval_{\symIntervalIdx}$ and we can continue processing the next smaller interval. Otherwise, if $\symRandomBit_\symIntervalIdx=1$, there is at least one active index, but we still need to find out which indices from $\symInterval_{\symIntervalIdx}$ are actually active.

As all indices $\symBucketIdx\in\symInterval_{\symIntervalIdx}$ have the same chance that their associated random value $\symRndVal_\symBucketIdx$ is the smallest among them, the largest active index in $\symInterval_{\symIntervalIdx}$ can simply be generated by drawing a uniform random value $\symRandomInt_\symIntervalIdx$ from $\symInterval_{\symIntervalIdx}$.
It is possible that $\symInterval_{\symIntervalIdx}$ contains further (smaller) active indices in the interval.
As the next smaller active index (not necessarily in $\symInterval_\symIntervalIdx$) is uniformly distributed over $\llbracket \symRandomInt_\symIntervalIdx\rrbracket$, we sample a uniform random value $\symRandomSubInt_{\symIntervalIdx,0}$ from $\llbracket\symRandomInt_\symIntervalIdx\rrbracket$ analogous to \eqref{equ:recursion_backwards}. If $\symRandomSubInt_{\symIntervalIdx,0}\notin\symInterval_{\symIntervalIdx}$, $\symRandomSubInt_{\symIntervalIdx,0}$ is rejected as active index and the process is continued with the next smaller interval $\symInterval_{\symIntervalIdx-1}$. Otherwise, $\symRandomSubInt_{\symIntervalIdx,0}$ is the next smaller active index within $\symInterval_{\symIntervalIdx}$, and we iteratively generate further random values according to
\begin{equation}
  \label{equ:recursion_z}
  \symRandomSubInt_{\symIntervalIdx,\symIntervalSubIdx}\sim\symUniform(\llbracket \symRandomSubInt_{\symIntervalIdx,\symIntervalSubIdx-1}\rrbracket)
  \qquad \text{with $\symIntervalSubIdx \geq 1$}
\end{equation}
as long as they are within $\symInterval_{\symIntervalIdx}$, to get all active indices of that interval.

The whole procedure to generate active indices in descending order, which is statistically equivalent to \cref{alg:jumping_backwards}, is summarized in \cref{alg:jumping_backwards_efficient}.
While $\symRandomBit_\symIntervalIdx$ and $\symRandomInt_\symIntervalIdx$ are already very convenient to sample, $\symRandomSubInt_{\symIntervalIdx,\symIntervalSubIdx}$ needs to be sampled from a non-power-of-two interval $\llbracket\symBucketIdx\rrbracket$.
One possibility would be to use Lemire's method \cite{Lemire2019}.
A simpler and also unbiased approach that fits better in our case is to sample from the smallest enclosing power-of-two interval, which is by definition $\llbracket2^{\symIntervalIdx+1}\rrbracket$ for $\symInterval_\symIntervalIdx$, according to
\begin{equation}
  \label{equ:comp_z_option_2}
  \symRandomSubInt_{\symIntervalIdx,\symIntervalSubIdx}\sim\symUniform(\llbracket2^{\symIntervalIdx+1}\rrbracket)
  \qquad \text{with $\symIntervalSubIdx \geq 1$}.
\end{equation}
As the sequence $(\symRandomSubInt_{\symIntervalIdx,\symIntervalSubIdx})_{\symIntervalSubIdx\geq 0}$ is no longer strictly monotonically decreasing, we must ignore values that are not smaller than all previous values to ensure equivalent behavior, when replacing \eqref{equ:recursion_z} in the inner loop of \cref{alg:jumping_backwards_efficient}, labeled as option 1, by \eqref{equ:comp_z_option_2}, labeled as option 2.
Fortunately, the inner loop takes care of this without any further modifications. If there is a value that fulfills one of the two stop conditions, but there was already another smaller or equal value before, this other value would have stopped the loop earlier.
We continue with option 2, as all required random values are then either Boolean values or values from a power-of-two interval and can therefore be generated easily and efficiently by taking a corresponding number of random bits.

\myAlg{
  \caption{Consistent hashing based on a more efficient generation of active indices in reverse order, $\symNumBuckets\leq 2^\symIntervalIdxMax$.}
  \label{alg:jumping_backwards_efficient}
  \SetKwFunction{FuncJumpingBackwardsImproved}{JumpingBackwardsImproved}
  \Fn{\FuncJumpingBackwardsImproved{$\symKey$, $\symNumBuckets$}}{
    $\symPRG\gets$ initialize \acs*{PRG} with seed $\symKey$ \;
    \For(\Comment*[f]{process intervals $\symInterval_{\symIntervalIdxMax-1}, \symInterval_{\symIntervalIdxMax-2},\ldots,\symInterval_{0}$ in descending order}){$\symIntervalIdx\gets\symIntervalIdxMax-1$ \KwTo $0$}{
      $\symRandomBit_\symIntervalIdx\gets\symPRG[\symUniform(\lbrace 0, 1\rbrace)]$\Comment*[r]{generate a random bit}
      \lIf(\Comment*[f]{continue with next interval if $\symInterval_{\symIntervalIdx}$ does not contain any active indices}){$\symRandomBit_\symIntervalIdx = 0$}{\Continue}
      $\symRandomInt_\symIntervalIdx\gets 2^{\symIntervalIdx} + \symPRG[\symUniform(\llbracket2^{\symIntervalIdx}\rrbracket)]$\Comment*[r]{sample largest active index $\symRandomInt_\symIntervalIdx$ from $\symInterval_\symIntervalIdx$}
      $\symBucketIdx\gets\symRandomInt_\symIntervalIdx$\;
      \For{$\symIntervalSubIdx\gets 0$ \KwTo $\infty$}{
        \lIf(\Comment*[f]{the first active index in $\symInterval_{\symIntervalIdx}$ smaller than $\symNumBuckets$ is returned}){$\symBucketIdx < \symNumBuckets$}{\Return{$\symBucketIdx$}}
        $\symRandomSubInt_{\symIntervalIdx,\symIntervalSubIdx}\gets\symPRG[\symUniform(\llbracket\symBucketIdx\rrbracket)]$\Comment*[r]{option 1, compare \eqref{equ:recursion_z}}
        $\symRandomSubInt_{\symIntervalIdx,\symIntervalSubIdx}\gets\symPRG[\symUniform(\llbracket 2^{\symIntervalIdx+1}\rrbracket)]$\Comment*[r]{option 2, compare \eqref{equ:comp_z_option_2}}
        $\symBucketIdx\gets \symRandomSubInt_{\symIntervalIdx,\symIntervalSubIdx}$\;
        \lIf(\Comment*[f]{stop if there are no further active indices in $\symInterval_{\symIntervalIdx}$}){$\symBucketIdx < 2^{\symIntervalIdx}$}{\Break}
      }
    }
    \Return{$0$}\;
  }
}

\subsection{Saving Random Values}
\label{sec:saving_random_values}
\cref{alg:jumping_backwards_efficient} requires many random values. For every interval $\symInterval_\symIntervalIdx$ a bit $\symRandomBit_\symIntervalIdx\in\lbrace0,1\rbrace$, an integer $\symRandomInt_\symIntervalIdx\in\symInterval_\symIntervalIdx$, and an integer sequence $(\symRandomSubInt_{\symIntervalIdx,\symIntervalSubIdx})_{\symIntervalSubIdx\geq 0}$ with $\symRandomSubInt_{\symIntervalIdx,\symIntervalSubIdx}\in\llbracket 2^{\symIntervalIdx+1}\rrbracket$ need to be sampled when using option 2. The sequences $(\symRandomSubInt_{\symIntervalIdx,\symIntervalSubIdx})_{\symIntervalSubIdx\geq 0}$ are finite, as it is sufficient to generate values up to the first value that is smaller than $2^\symIntervalIdx$ and that causes leaving the inner loop of \cref{alg:jumping_backwards_efficient} in any case.
We are free to change the order in which those independent random values and sequences are generated using the same \ac{PRG} without changing the statistical properties of \cref{alg:jumping_backwards_efficient}. However, the order must be well-defined and not depend on $\symNumBuckets$ in order to preserve monotonicity.

As first optimization, we generate all $\symIntervalIdxMax$ random bits $\symRandomBit_\symIntervalIdx$ at once by drawing a random $\symIntervalIdxMax$-bit integer $\symRandomBitAll = \langle\symRandomBit_{\symIntervalIdxMax-1}\symRandomBit_{\symIntervalIdxMax-2}\ldots \symRandomBit_0\rangle_2 \in \llbracket 2^\symIntervalIdxMax\rrbracket$. Knowing all those bits at once allows quick identification of intervals that may contain active indices that are relevant for a given $\symNumBuckets$.
The indices of all set bits of $\symRandomBitAll$ can be computed in descending order according to the recursion
\begin{equation}
  \label{equ:mi}
  \symNonZero_{\symNonZeroIndex+1} = \lfloor\log_2(\symRandomBitAll \bmod 2^{\symNonZero_{\symNonZeroIndex}})\rfloor
\end{equation}
as long as the argument of the binary logarithm is nonzero. If it is zero there are no further intervals $\symInterval_\symIntervalIdx$ with active indices.
As only intervals $\symInterval_\symIntervalIdx$ are relevant that have the corresponding bit $\symRandomBit_\symIntervalIdx$ set and that have an overlap with $\llbracket \symNumBuckets\rrbracket$, we can start our recursion with interval index
\begin{equation}
  \label{equ:m_0}
  \symNonZero_0 = \lfloor \log_2(\symNumBuckets-1)\rfloor+1
  \qquad \text{for $\symNumBuckets\geq 2$}
\end{equation}
which is the solution of the inequality
\begin{equation*}
  2^{\symNonZero_0-1} < \symNumBuckets \leq 2^{\symNonZero_0}.
\end{equation*}
Thus, $\symInterval_{\symNonZero_0}$ is the smallest interval without overlap with $\llbracket \symNumBuckets\rrbracket$.

Among the set of intervals $\lbrace \symInterval_{\symIntervalIdx}\rbrace_{\symIntervalIdx\in\llbracket\symIntervalIdxMax\rrbracket}$, at most two need to be processed\chreplaced{, because}{ as} $\symInterval_{\symIntervalIdxOne}$ and $\symInterval_{\symIntervalIdxTwo}$ (if they exist) are the largest two intervals that contain at least one active index and overlap with $\llbracket \symNumBuckets\rrbracket$. As $\symInterval_{\symIntervalIdxTwo}\subseteq\llbracket \symNumBuckets\rrbracket$ by definition, $\symInterval_{\symIntervalIdxTwo}$ must contain an active index that is also smaller than $\symNumBuckets$. Therefore, the active indices of smaller intervals $\symInterval_{\symIntervalIdx}$ with $\symIntervalIdx<\symIntervalIdxTwo$ are not relevant. Furthermore, within $\symInterval_{\symIntervalIdxTwo}$ just the largest active index is needed, which means that the random sequence
$(\symRandomSubInt_{\symIntervalIdxTwo,\symIntervalSubIdx})_{\symIntervalSubIdx\geq 0}$ is never used.
The only sequence that may be required for generating the largest active index smaller than $\symNumBuckets$ is $(\symRandomSubInt_{\symIntervalIdxOne,\symIntervalSubIdx})_{\symIntervalSubIdx\geq 0}$ as the active indices of $\symInterval_\symIntervalIdxOne$ are not necessarily smaller than $\symNumBuckets$.

Since no more than one of the sequences $(\symRandomSubInt_{0,\symIntervalSubIdx})_{\symIntervalSubIdx\geq 0}$, $(\symRandomSubInt_{1,\symIntervalSubIdx})_{\symIntervalSubIdx\geq 0},\ldots,(\symRandomSubInt_{\symIntervalIdxMax-1,\symIntervalSubIdx})_{\symIntervalSubIdx\geq 0}$ is used at the same time, it is not necessary that those sequences are statistically independent to ensure uniformity.
Therefore, we are allowed to generate a single sequence $(\symRandomSubIntAll_\symIntervalSubIdx)_{\symIntervalSubIdx\geq 0}$ with $\symRandomSubIntAll_\symIntervalSubIdx\in\llbracket 2^{\symIntervalIdxMax}\rrbracket$ instead and set
\begin{equation}
  \label{equ:zmv}
  \symRandomSubInt_{\symIntervalIdx,\symIntervalSubIdx} = \symRandomSubIntAll_\symIntervalSubIdx \bmod 2^{\symIntervalIdx+1}.
\end{equation}
It should be noted that with this modification, the generation of active indices is no longer statistically equivalent to \eqref{equ:basic_consistent}. However, uniformity is preserved as the consumed random values are still independent. Monotonicity is also still ensured, as the active indices are still generated in descending order.

We can also reduce the randomness needed for the values $\symRandomInt_\symIntervalIdx$.
As only active indices of at most the two intervals $\symInterval_\symIntervalIdxOne$ and $\symInterval_\symIntervalIdxTwo$ are relevant, only the two random integers $\symRandomInt_\symIntervalIdxOne$ and $\symRandomInt_\symIntervalIdxTwo$ are needed at the same time. Thus, using the fact that $1 + \sum_{\symSomeIdx=0}^\symIntervalIdxTwo \symRandomBit_\symSomeIdx = \sum_{\symSomeIdx=0}^\symIntervalIdxOne \symRandomBit_\symSomeIdx$, we can set
\begin{equation}
  \label{equ:y_m}
  \symRandomInt_\symIntervalIdx = 2^\symIntervalIdx + (\symRandomIntAll_\symSelectIdx \bmod 2^\symIntervalIdx)
  \qquad \text{with $\textstyle\symSelectIdx=(\sum_{\symSomeIdx=0}^\symIntervalIdx \symRandomBit_\symSomeIdx ) \bmod 2$}
\end{equation}
and with $\symRandomIntAll_0, \symRandomIntAll_1 \in\llbracket 2^{\symIntervalIdxMax}\rrbracket$ being two independent random values, as this is sufficient for the independence of $\symRandomInt_\symIntervalIdxOne$ and $\symRandomInt_\symIntervalIdxTwo$.

If we look at the random variables $\symRandomIntAll_0$ and $\symRandomIntAll_1$, the $\symIntervalIdxOne$ least significant bits are used from one and the least significant $\symIntervalIdxTwo$ bits from the other. As a consequence, the most significant $\symIntervalIdxMax-\symIntervalIdxTwo$ bits of either $\symRandomIntAll_0$ or $\symRandomIntAll_1$ are still unused. After combining both using the XOR-operation, the most significant $\symIntervalIdxMax-\symIntervalIdxTwo$ bits of $\symRandomIntAll_0\bitwiseXor\symRandomIntAll_1$ are therefore statistically independent of $\symRandomInt_\symIntervalIdxOne$ and $\symRandomInt_\symIntervalIdxTwo$.
This allows us to use $\symRandomIntAll_0$ and $\symRandomIntAll_1$ also for
$\symRandomBitAll$ by setting
\begin{equation}
  \label{equ:u_xor}
  \symRandomBitAll = \symRandomIntAll_0\bitwiseXor\symRandomIntAll_1,
\end{equation}
because we only need the random bits
$\symRandomBit_{\symIntervalIdxMax-1},\ldots, \symRandomBit_{\symIntervalIdxTwo}$ to determine $\symIntervalIdxOne$ and $\symIntervalIdxTwo$ and therefore never consume more than the leading $\symIntervalIdxMax-\symIntervalIdxTwo$ bits of $\symRandomBitAll$.

In summary, it is sufficient to generate $\symRandomIntAll_0$ and $\symRandomIntAll_1$ followed by the sequence $(\symRandomSubIntAll_\symIntervalSubIdx)_{\symIntervalSubIdx\geq 0}$ to generate all needed random values without breaking uniformity and monotonicity.
Therefore, all these values can be taken from a single random sequence which allows to use a standard \ac{PRG}.

\subsection{JumpBackHash}

\myAlg{
  \caption{JumpBackHash, $\symNumBuckets\leq 2^\symIntervalIdxMax$.}
  \label{alg:jump_back_hash}
  \SetKwFunction{FuncJumpBackHash}{JumpBackHash}
  \Fn{\FuncJumpBackHash{$\symKey$, $\symNumBuckets$}}{
    \lIf{$\symNumBuckets \leq 1$}{\Return{0}}
    $\symPRG\gets$ initialize \acs*{PRG} with seed $\symKey$ \;
    $\symRandomIntAll_0\gets \symPRG[\symUniform(\llbracket2^\symIntervalIdxMax\rrbracket)]$\;
    $\symRandomIntAll_1\gets \symPRG[\symUniform(\llbracket2^\symIntervalIdxMax\rrbracket)]$\;
    $\symRandomBitAll \gets \symRandomIntAll_0\bitwiseXor \symRandomIntAll_1$\Comment*[r]{compare \eqref{equ:u_xor}}
    $\symIntervalIdx \gets \lfloor \log_2(\symNumBuckets-1)\rfloor+1$\Comment*[r]{equals $\symNonZero_0$, see \eqref{equ:m_0}}
    $\symSubBits\gets
      \symRandomBitAll\bmod 2^{\symIntervalIdx}$\;
    \While(\Comment*[f]{iterate over intervals $\symInterval_{\symIntervalIdx_1},\symInterval_{\symIntervalIdx_2},\ldots$ that contain at least one active index}){$\symSubBits\neq 0$}{
      $\symIntervalIdx\gets \lfloor\log_2 \symSubBits\rfloor$\Comment*[r]{equals
        $\symNonZero_\symNonZeroIndex$ in $\symNonZeroIndex$-th iteration, compare \eqref{equ:mi}}
      $\symSelectIdx \gets \symBitCount(\symSubBits) \bmod 2$\Comment*[r]{determine parity of set bits of $\symSubBits$}
      $\symBucketIdx\gets 2^{\symIntervalIdx} + (\symRandomIntAll_\symSelectIdx \bmod 2^\symIntervalIdx)$\Comment*[r]{corresponds to  $\symRandomInt_\symIntervalIdx$, compare \eqref{equ:y_m}}
      \For(\Comment*[f]{$\symIntervalSubIdx$ could be eliminated but is kept for better comparability with \cref{alg:jumping_backwards_efficient}}){$\symIntervalSubIdx\gets 0$ \KwTo $\infty$}{
        \lIf(\Comment*[f]{condition is always satisfied \chadded{at} latest in the second while-iteration}){$\symBucketIdx < \symNumBuckets$}{\Return{$\symBucketIdx$}}
        $\symRandomSubIntAll_{\symIntervalSubIdx}\gets\symPRG[\symUniform(\llbracket 2^{\symIntervalIdxMax}\rrbracket)]$\;
        $\symBucketIdx\gets \symRandomSubIntAll_{\symIntervalSubIdx}\bmod2^{\symIntervalIdx+1}$\Comment*[r]{corresponds to $\symRandomSubInt_{\symIntervalIdx,\symIntervalSubIdx}$, compare \eqref{equ:zmv}}
        \lIf(\Comment*[f]{stop if there are no further active indices in $\symInterval_{\symIntervalIdx}$}){$\symBucketIdx < 2^{\symIntervalIdx}$}{\Break}
      }
      $\symSubBits\gets\symSubBits\bitwiseXor 2^\symIntervalIdx$\Comment*[r]{has the same effect as $\symSubBits\gets\symSubBits\bmod 2^{\symIntervalIdx}$}
    }
    \Return{$0$}\;
  }
}

Applying all these optimizations to \cref{alg:jumping_backwards_efficient} with option 2 finally leads to JumpBackHash as shown in \cref{alg:jump_back_hash}. The variables $\symRandomBitAll$ and $\symRandomSubIntAll_{\symIntervalSubIdx}$ as well as the loop variable $\symIntervalSubIdx$ could be eliminated, but are kept for better comprehensibility and better comparability with \cref{alg:jumping_backwards_efficient}. The while loop iterates over the intervals $\symInterval_{\symIntervalIdx_1},\symInterval_{\symIntervalIdx_2},\ldots$ according to \eqref{equ:mi}.
Since we know that $\symInterval_{\symIntervalIdx_2}$, if it exists, must contain an active index smaller than $\symNumBuckets$, the condition for the return statement within the inner loop for $\symInterval_{\symIntervalIdx_2}$ is always fulfilled. This means that there will never be more than two iterations of the while loop.

\cref{alg:jump_back_hash_java} shows how the algorithm can be implemented with a few lines of code in Java.
\chadded{As most state-of-the-art hash functions produce 64-bit values \cite{Yi,Vaneev, Peters}, we assumed the key to be of type \lstinline[keywordstyle=\color{black}]{long}. Furthermore, the type for the number of buckets was chosen to be an \lstinline[keywordstyle=\color{black}]{int} which is large enough for most use cases and ensures uniformity as it is always much smaller than $2^{64}$ (c.f. \cref{sec:consistent_hash_function}).}
Expressions of kind $\lfloor\log_2 \symSubBits\rfloor$ can be replaced by \lstinline{31-numberOfLeadingZeros(u)} if $\symSubBits$ is a positive 32-bit integer. Furthermore, bit-shift operations are well-defined in Java. For example, shifting by $-29$ bits is the same as shifting by 3 bits. Therefore, shifting by \lstinline{31-numberOfLeadingZeros(u)} is the same as shifting by \lstinline{-1-numberOfLeadingZeros(u)} or by \lstinline{~numberOfLeadingZeros(u)} when using Java's bitwise complement operator.
The \lstinline{numberOfLeadingZeros} and \lstinline{bitCount} Java functions are intrinsic candidates. They are usually translated directly into specific CPU instructions and are therefore very fast.

As \acp{PRG} typically produce 64-bit values, we can use them to get two 32-bit random integers at once. Therefore, \cref{alg:jump_back_hash_java} uses the least and most significant 32 bits of the first 64-bit random value as $\symRandomIntAll_0$ and $\symRandomIntAll_1$, respectively. Also the inner while loop uses a 64-bit random value to always generate two elements of the sequence $(\symRandomSubIntAll_{\symIntervalSubIdx})_{\symIntervalSubIdx\geq 0}$ at once. If two random values are always generated at once, we refer to JumpBackHash* in the following.

\newsavebox{\codebox}
\begin{lrbox}{\codebox}
  \begin{lstlisting}
int jumpBackHash(long k, int n) {
  if (n <= 1) return 0;
  randomGenerator.resetWithSeed(k);
  long v = randomGenerator.nextLong();
  int u = (int) (v ^ (v >>> 32)) & (~0 >>> numberOfLeadingZeros(n - 1));
  while (u != 0) {
    int q = 1 << ~numberOfLeadingZeros(u); // q = 2^m
    int b = q + ((int) (v >>> (bitCount(u) << 5)) & (q - 1));
    while (true) {
      if (b < n) return b;
      long w = randomGenerator.nextLong();
      b = (int) w & ((q << 1) - 1);
      if (b < q) break;
      if (b < n) return b;
      b = (int) (w >>> 32) & ((q << 1) - 1);
      if (b < q) break;
    }
    u ^= q;
  }
  return 0;
}
\end{lstlisting}
\end{lrbox}
\myAlg{
  \caption{Java implementation of JumpBackHash using a 64-bit \acl*{PRG}.}
  \label{alg:jump_back_hash_java}
  \usebox{\codebox}
}

\subsection{Runtime Analysis}
\label{sec:runtime_analysis}
The runtime of \cref{alg:jump_back_hash} scales with the number of consumed random values. Therefore, we analyze the distribution of the number of consumed random values which corresponds to the number of \ac{PRG} calls. As the case $\symNumBuckets=1$ does not require any random values, we can focus on the case $\symNumBuckets\geq 2$.

First, we need two random values for $\symRandomIntAll_0$ and $\symRandomIntAll_1$ (compare \cref{sec:saving_random_values}). Further random values are only needed if the largest active index in $\symInterval_{\symIntervalIdx_1}$ is greater than or equal to $\symNumBuckets$. This can only happen, if $\symNumBuckets\in\symInterval_{\symIntervalIdx_1}$, and thus if $\symIntervalIdx_1 = \symIntervalIdx_0 -1$. The probability that $\symInterval_{\symIntervalIdx_0 -1}$ contains an active index at all is $\frac{1}{2}$. As the largest active index would be uniformly distributed over $\symInterval_{\symIntervalIdx_0 -1}$, the probability that $\symInterval_{\symIntervalIdx_0 -1}$ contains an active index that is larger than or equal to $\symNumBuckets$, or equivalently from $[\symNumBuckets, 2^{\symIntervalIdx_0})$,  is given by $\frac{1}{2}\cdot\frac{2^{\symIntervalIdx_0}-\symNumBuckets}{2^{\symIntervalIdx_0-1}} = 1-\symRatio^{-1}$
with
\begin{equation}
  \symRatio := \frac{2^{\symIntervalIdx_0}}{\symNumBuckets}
  =
  \frac{2^{\lfloor \log_2(\symNumBuckets-1)\rfloor+1}}{\symNumBuckets}
  \in[1,2).
\end{equation}
Only if this is the case we need to start generating the random sequence $(\symRandomSubIntAll_{\symIntervalSubIdx})_{\symIntervalSubIdx\geq 0}$ to find the largest active index within $\symInterval_{\symIntervalIdx_0 -1}$ that is actually smaller than $\symNumBuckets$. The inner loop is left, either via the break or the return statement, as soon as $\symRandomSubIntAll_{\symIntervalSubIdx} \bmod 2^{\symIntervalIdx_1+1} = \symRandomSubIntAll_{\symIntervalSubIdx} \bmod 2^{\symIntervalIdx_0} < \symNumBuckets$ which occurs with probability $\frac{\symNumBuckets}{2^{\symIntervalIdx_0}}=\symRatio^{-1}$. Therefore, the number of random values consumed within the inner loop follows a geometric distribution with positive support and a success probability of $\symRatio^{-1}$.

In summary, the number of random values $\symNumberRandomValuesJBH$ that need to be generated is distributed like
\begin{equation}
  \symNumberRandomValuesJBH
  \sim
  2 + \symBernoulli(1 - \symRatio^{-1})\cdot \symGeometric(\symRatio^{-1})
  \qquad \text{for $\symNumBuckets\geq2$}.
\end{equation}
Using the identities for the \chreplaced{expected value}{mean} and variance of the product of independent $\symSomeRandomVariateOne$ and $\symSomeRandomVariateTwo$ \cite{Goodman1960}
\begin{align*}
  \symExpectation(\symSomeRandomVariateOne \symSomeRandomVariateTwo) & = \symExpectation(\symSomeRandomVariateOne)\symExpectation(\symSomeRandomVariateTwo),
  \\
  \symVariance(\symSomeRandomVariateOne \symSomeRandomVariateTwo)    & = (\symExpectation(\symSomeRandomVariateOne ))^2 \symVariance(\symSomeRandomVariateTwo)+ (\symExpectation(\symSomeRandomVariateTwo ))^2 \symVariance(\symSomeRandomVariateOne) + \symVariance(\symSomeRandomVariateOne)\symVariance(\symSomeRandomVariateTwo)
\end{align*}
together with \eqref{equ:bernoulli_expectation}, \eqref{equ:bernoulli_variance}, and the known expressions for geometrically distributed random variables
\begin{align*}
  \symExpectation(\symGeometric(\symSuccessProbability)) & =\frac{1}{\symSuccessProbability},
  \\
  \symVariance(\symGeometric(\symSuccessProbability))    & =\frac{1-\symSuccessProbability}{\symSuccessProbability^2},
\end{align*}
the \chreplaced{expected value}{mean} and the variance of $\symNumberRandomValuesJBH$ can be derived as
\begin{alignat}{3}
  \label{equ:jph_mean}
  \symExpectation
   &
  (\symNumberRandomValuesJBH)
   &   &
  =
  1+ \symRatio
   &   &
  \in [2,3),
  \\
  \label{equ:jph_var}
  \symVariance
   &
  (\symNumberRandomValuesJBH)
   &   &
  =
  (\symRatio-1)\symRatio
   &   &
  \in[0,2).
\end{alignat}
Hence, on average, JumpBackHash requires less than 3 random values with a variance that is always less than 2.

For JumpBackHash*, when two random values are always generated at once as in \cref{alg:jump_back_hash_java},
the probability that \chadded{at least} one of both random values satisfies any stop condition within the inner loop is given by $1-(1-\symRatio^{-1})(1-\symRatio^{-1})=\symRatio^{-1}(2-\symRatio^{-1})$.
Therefore, as also just a single random value is needed at the beginning, the number of generated random values is distributed as
\begin{equation}
  \symNumberRandomValuesJBHStar
  \sim
  1 + \symBernoulli(1 - \symRatio^{-1})\cdot \symGeometric(\symRatio^{-1}(2-\symRatio^{-1}))
  \qquad \text{for $\symNumBuckets\geq2$}
\end{equation}
with \chreplaced{expected value}{mean} and variance given by
\begin{alignat}{3}
  \label{equ:jph_star_mean}
  \symExpectation
   &
  (\symNumberRandomValuesJBHStar)
   &   &
  =
  1 + \frac{(\symRatio-1)\symRatio}{2\symRatio - 1}
   &   &
  \in
  \textstyle[1,{\frac{5}{3}}),
  \\
  \label{equ:jph_star_var}
  \symVariance
   &
  (\symNumberRandomValuesJBHStar)
   &   &
  =
  \frac{\symRatio(\symRatio-1)(\symRatio^2-\symRatio+1)}{(2\symRatio - 1)^2}
   &   &
  \in
  \textstyle[0,{\frac{2}{3}}).
\end{alignat}
Hence, \cref{alg:jump_back_hash_java} consumes on average at most $\frac{5}{3}\approx 1.667$ 64-bit random integers.

$\symRatio$ is minimized and equal to 1 if $\symNumBuckets$ is exactly a power of two. The maxima of $\symRatio$ occur whenever the number of buckets are powers of two incremented by 1, $\symNumBuckets=2^\symSomeIdx+1$ and approach 2 as $\symNumBuckets\rightarrow\infty$. This leads to a sawtooth pattern on the log-scale. As the expressions \eqref{equ:jph_mean}, \eqref{equ:jph_var}, \eqref{equ:jph_star_mean}, and \eqref{equ:jph_star_var} are all monotonically increasing functions of $\symRatio$ over $[1,2)$, they replicate this sawtooth pattern. In contrast to JumpHash, where not only the \chreplaced{expected value}{mean} \eqref{equ:jh_mean} but also the variance \eqref{equ:jh_var} increases logarithmically, both are bounded in case of JumpBackHash.

\section{Experimental Verification}
To validate JumpBackHash, we have performed a series of tests. We used the SplitMix random generator \cite{Steele2014} which produces a sequence of 64-bit random integers, is known to have good statistical properties \cite{TestingRNG}, and is also available in the Java standard library as SplittableRandom. The source code to run all tests and to reproduce the presented results has been published at
\url{https://github.com/dynatrace-research/jumpbackhash-paper}.
The tested JumpBackHash implementation is equivalent to the one included in our Hash4j Java open source library since version 0.17.0 \cite{Hash4j}.

\begin{figure}
  \centering
  \includegraphics[width=0.75\linewidth]{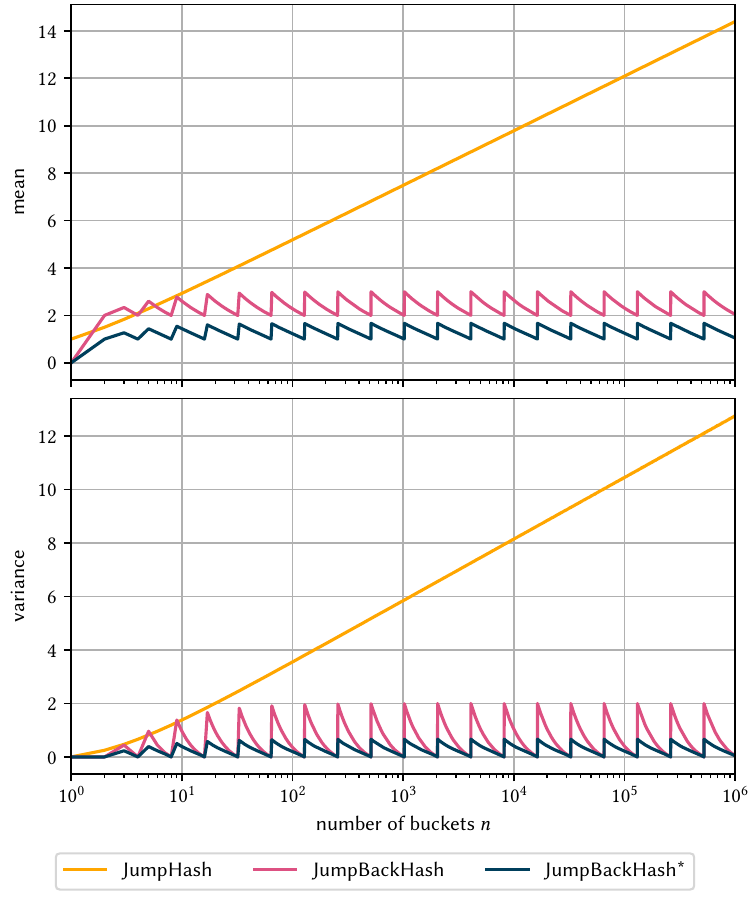}
  \caption{\boldmath The measured mean and variance of the number of consumed random values for JumpHash and JumpBackHash. JumpBackHash* corresponds to the case when a single random value is split into two as in \cref{alg:jump_back_hash_java}.}
  \label{fig:time_complexity}
\end{figure}

\subsection{Consistency}
To verify consistency we must verify monotonicity and uniformity.
Monotonicity is straightforward to test. We did \num{10000} simulation runs, in each of which we used a random key and computed the assignments for all $\symNumBuckets\in\lbrace 1,2,\ldots,\num{10000}\rbrace$ and checked, whenever the reassignment changed after an increment of $\symNumBuckets$, that the new assignment corresponds to the new bucket according to \eqref{equ:def_mononicity}.

Testing uniformity is more sophisticated as it requires statistical means. Our strategy was to sample many bucket indices by calling the consistent hash function for many different keys. In our test we always used 1 million randomly generated keys. If uniformity holds, we expect that the bucket index distribution is uniform. For small numbers of buckets, as long as it is much smaller than the number of sampled keys, this can be verified by applying the G-test \cite{McDonald2014}. In this way we checked the uniformity for all $\symNumBuckets\in \lbrace 1,2,\ldots,\num{1000}\rbrace$.
Additionally, we also investigated if uniformity holds for large $\symNumBuckets$ close to $2^{31}-1$ which is the largest possible positive 32-bit integer value in Java. In particular, we have tested the values
$\symNumBuckets\in\lbrace 2^{31}-1, 2^{31}-2, 3\cdot 2^{29}, 2^{30}+1, 2^{30}, 2^{30}-1, 3\cdot 2^{28},2^{29}+1, 2^{29}, 2^{29}-1, 3\cdot 2^{27},2^{28}+1, 2^{28}, 2^{28}-1\rbrace$ using the Kolmogorov-Smirnov test \cite{Kanji2006}, as the number of occurrences per possible outcome would be too low for a valid application of the G-test.

\subsection{Random Value Consumption}

We also checked the analytical expressions for the number of consumed random values, which is a measure of the runtime. For this purpose, we ran simulations for different $\symNumBuckets$. To cover many orders of magnitude, we picked values for $\symNumBuckets$ according to the recursion
\begin{equation}
  \symNumBuckets_{\symSomeIdx+1} = \lfloor0.999\cdot\symNumBuckets_{\symSomeIdx} \rfloor\qquad \text{with $\symNumBuckets_{0}=10^6$}.
\end{equation}
This results in 7482 positive integer values that are roughly uniformly distributed on the logarithmic scale over the range $[1,10^6]$.

For each value of $\symNumBuckets$ we computed the bucket index for 10 million random keys and counted how often the pseudorandom generator was called. The corresponding sample mean and variance are shown in \cref{fig:time_complexity} for JumpHash and JumpBackHash. We also considered JumpBackHash* which splits a random value into two and which corresponds to \cref{alg:jump_back_hash_java}.
All simulated data points for the mean did not differ from the theoretical prediction by more than 0.0036 in absolute values. Likewise, the observed difference between the empirical and the theoretical variance was always less than 0.025. We have therefore omitted plotting the theoretical curves as they look identical and overlap perfectly, which also confirms the correctness of our theoretical analysis.

\subsection{Performance Measurements}

\begin{figure}
  \centering
  \includegraphics[width=0.75\linewidth]{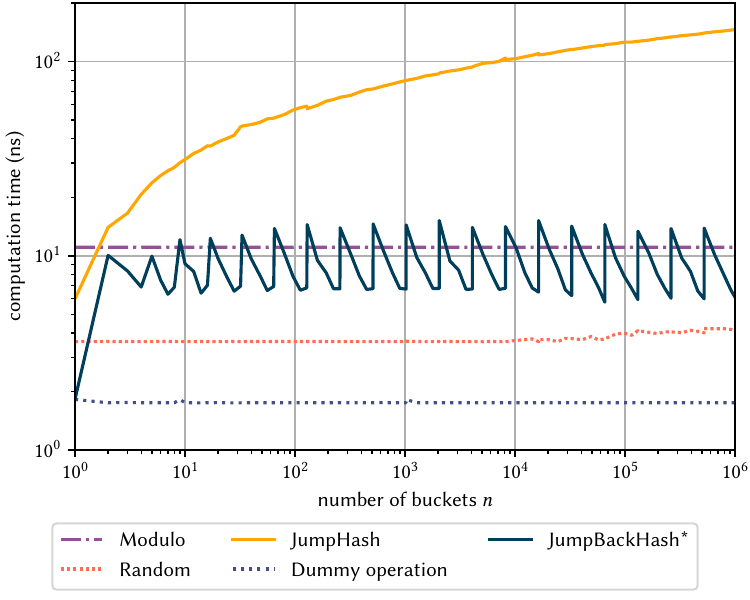}
  \caption{\boldmath Benchmark results on an Amazon EC2 c5.metal instance with Intel Xeon Platinum 8275CL CPU and processor P-state set to 1. Unlike \cref{fig:time_complexity}, a logarithmic scale is used for the y-axis.}
  \label{fig:benchmark}
\end{figure}

We evaluated the performance of JumpBackHash on an Amazon EC2 c5.metal instance with Intel Xeon Platinum 8275CL CPU. To reduce the impact of variable processor frequencies, Turbo Boost was disabled by setting the processor P-state to 1 \cite{AmazonPState}. The benchmarks were implemented using the \ac{JMH} toolkit and were executed using OpenJDK 21.0.2.

The number of buckets was varied from 1 to \num{e6}. We included all values that can be represented as $2^\symSomeIdx, 2^\symSomeIdx+1, \lfloor 2^\symSomeIdx(1+\frac{1}{4})\rfloor, \lfloor 2^\symSomeIdx(1+\frac{1}{2})\rfloor, \lfloor 2^\symSomeIdx(1+\frac{3}{4})\rfloor$ with some integer $\symSomeIdx$. The first two correspond to the best and the worst cases for JumpBackHash according to our theoretical analysis.

The benchmark results for JumpHash (\cref{alg:jumphash}) and JumpBackHash* (\cref{alg:jump_back_hash_java}) are shown in \cref{fig:benchmark}. As reference, we also included non-consistent approaches based on the modulo-operation \eqref{equ:modulo_pattern} and a pseudorandom assignment using Lemire's algorithm \cite{Lemire2019}. Furthermore, we added a dummy test that just consumed the random keys to make sure that the input is not the bottleneck. To allow a fairer comparison, we used SplitMix \cite{Steele2014} for generating random values for all approaches that require a \ac{PRG}.

The results agree with the theoretical runtime analysis and show that JumpBackHash is always faster than JumpHash. Interestingly, the runtime of JumpBackHash is comparable or even faster than the modulo-based approach. The reason is that the modulo operation is a quite expensive operation as it involves an integer division. Therefore, if only uniformity but no monotonicity is needed, it would be much better to use the pseudorandom approach.

\section{Future Work}
In our tests we used the SplitMix random generator \cite{Steele2014} which is already quite fast and known to have good quality \cite{TestingRNG}.
However, there might be even faster random generators
\chreplaced{
  that have sufficient statistical quality for generating the short sequences of random values as needed by JumpBackHash. For example, \acp{LCG} \cite{LEcuyer1999} could be used. Even though their direct application could be problematic because the least significant bits of \acp{LCG} are known to have poor quality, the most significant bits still could be good enough. Linear xorshift random generators \cite{Marsaglia2003, Blackman2021} are a second interesting class of potentially faster generators.
}{
  such as linear congruential random generators \cite{LEcuyer1999} that have sufficient statistical quality for JumpBackHash.
}

Another idea, which also needs further investigation, is to use the hashed key directly as the first random value. In this way, one random value computation can be saved and the random generator only needs to be initialized with the key as seed when entering the inner loop. First experiments showed that our statistical tests were still successful when using the SplitMix generator. Of course, this can only work for \acp{PRG} where the random values are sufficiently independent of the seed.

\section{Conclusion}
We have introduced JumpBackHash, a new consistent hash algorithm for distributing keys to buckets.
It can be implemented with a few lines of code and requires only a standard random generator as found in standard libraries. As theoretically shown and experimentally confirmed, JumpBackHash has an expected constant runtime and is significantly faster than JumpHash. Its speed also allows safely replacing modulo-based assignments to improve system stability during changes by reducing the probability of spikes in resource consumption.

\bibliographystyle{ACM-Reference-Format}
\bibliography{bibliography}

\end{document}